\documentclass[11pt]{article}
\usepackage[top=1in,bottom=1in,right=1in,left=1in]{geometry}

\usepackage{float}
\usepackage{placeins}
\usepackage{pdflscape}

\usepackage{amsmath,amssymb,amsthm,amsfonts,bbold,xfrac}   
\usepackage{graphicx}   
\usepackage{pstricks}   
\usepackage{hyperref}   
\usepackage{listings}
\usepackage{xcolor}
\usepackage{color}

\usepackage{mathtools}      

\usepackage{mathrsfs}
\usepackage{sidecap}

\usepackage[maxbibnames=99]{biblatex}
\graphicspath{{./}{figures/}}

\renewbibmacro*{volume+number+eid}{
  \printfield{volume}
  \setunit*{\addnbspace}
  \printfield{number}
  \setunit{\addcomma\space}
  \printfield{eid}}
\DeclareFieldFormat[article]{number}{\mkbibparens{#1}}

\begin{document}

\title{Solving 3D Magnetostatics with RBF-FD:\\ 
Applications to the Solar Corona}

\author{Nathaniel H. Mathews \\
Dept. of Applied Mathematics, University of Colorado  \\
Boulder, CO 80309 USA\\
nathaniel.mathews@colorado.edu\footnote{Corresponding author}\\
\\
Natasha Flyer \\
Flyer Research LLC\\
Dept. of Applied Mathematics, University of Colorado  \\
Boulder, CO 80309 USA \\
natasha.flyer@colorado.edu\\
\\
Sarah E. Gibson \\
High Altitude Observatory\\
National Center for Atmospheric Research\\
Boulder, CO 80305 USA\\
sgibson@ucar.edu \\
}

\maketitle

\begin{abstract}
We present a novel magnetohydrostatic numerical model that solves directly for the force-balanced magnetic field in the solar corona. This model is constructed with Radial Basis Function Finite Differences (RBF-FD), specifically 3D polyharmonic splines plus polynomials, as the core discretization. This set of PDEs is particularly difficult to solve since in the limit of the forcing going to zero it becomes ill-posed with a multitude of solutions. For the forcing equal to zero there are no numerically tractable solutions. For finite forcing, the ability to converge onto a physically viable solution is delicate as will be demonstrated. The static force-balance equations are of a hyperbolic nature, in that information of the magnetic field travels along characteristic surfaces, yet they require an elliptic type solver approach for a sparse overdetermined ill-conditioned system. As an example,  we reconstruct a highly nonlinear analytic model designed to represent long-lived magnetic structures observed in the solar corona.
\end{abstract}

\section{Introduction}

The 3D forced magnetohydrostatics (MHS) equations constitute a very challenging set of partial differential equations (PDEs) with unusual properties that can cause difficulties:
\begin{itemize}
    \item In the limit of the forcing gong to zero, the PDEs become ill-posed in the sense that the number of continuous solutions goes to a set of measure zero. For the zero forcing case, physically viable solutions contain discontinuities \cite{Low_2007}.
    \item For the MHS equations with non-zero plasma forcing, the size and uniqueness of the solution space is highly dependent on the magnitude and type of forcing \cite{mathews20}.
    \item Solving for highly nonlinear solutions is delicate and can only be achieved by increments in resolution when applying a Newton solver approach to these hyberbolic-type PDEs.
    \item An iterative elliptic solver is needed, as the equations are not time-dependent but static. However, they are not elliptic PDEs in a classical definition, as the highest derivatives are first derivatives, giving them a hyberbolic nature in the sense that initial data does travel along characteristic surfaces.
\end{itemize}

For readers more familiar with classical fluid mechanics, a somewhat related situation occurs for steady flows. In the limit of vanishing viscosity, the Navier-Stokes equations lead to a fundamentally different set of solutions than the zero viscosity case, known as the Euler solutions which generally contain discontinuities. The infinite domain cases of 2-D flows past a cylinder and 3-D flows past a sphere are discussed in \cite{FE14}.

Due to the itemized issues above, a direct approach to solving for the static magnetic field, i.e. the Lorentz force balancing the plasma pressure (see Equation (1) directly below), has not to the authors' knowledge been published in the literature. Two common approaches are 1) magnetic relaxation techniques which often take millions of iterations to converge \cite{wiegelmann_neukirch06, zhu18} and 2) decomposing the magnetic field into components, such as the Toroidal and Poloidal fields or in terms of Euler potentials \footnote{Euler potentials represent the magnetic field as the cross product of two conservative fields}. However, Euler potentials break down at null points, and the Toroidal/Poloidal decomposition is generally nonunique. With regard to MHS, the latter technique often can only be applied to a linear subspace of MHS solutions \cite{MacTaggart13}. The presented solver does not have any of the above limitations.

The forced magnetohydrostatic equations are
\begin{equation}
\begin{split}
    \left(\nabla\times\mathbf{B}\right)\times\mathbf{B} &= \mathbf{F}\\
    \nabla\cdot\mathbf{B} &= 0
\end{split}
\label{eq_mhs}
\end{equation}
where $\mathbf{F}$ represents a known force exerted by the plasma medium on the magnetic field $\mathbf{B}$. This hydrostatic force is usually expressed in the literature as $\mathbf{F} = \nabla P + \rho g\hat{z}$ with $P$ the pressure, $\rho$ the density, $g$ the gravitational acceleration and $\hat{z}$ the vertical unit vector. Both the divergence and curl are 3D operators. 

The paper is organized as follows. Section 2 gives a brief introduction to the Radial Basis Function-generated Finite Differences (RBF-FD) method with polyharmonic splines plus polynomials. Section 3 shows the RBF-FD setup for the MHS equations. Section 4 provides a overview of the algorithm. It discusses how it is applied to the two-stage numerical solver. In the first stage, we execute a Quasi-Newton optimization with the analytic Jacobian to attain a force-balancing magnetic field. This is performed using a preconditioned Least-Squares (LSQR) iterative solver at each Newton step. In the second stage, we remove the remaining divergence from the magnetic field in a manner that minimally perturbs the force-balance obtained by the Newton approach. We demonstrate the performance of the solver against an analytic model that is often used to represent both long-lived and eruptive magnetic structures in the solar corona. 

\section{Brief Introduction to RBF-FD}

Radial Basis Function-generated Finite Differences (RBF-FD) can be considered a generalization of classical Finite Differences (FD) \cite{Shu2003,Tolstykh2003,Wright2006} to arbitrary node layouts. As in FD, RBF-FD approximates a linear differential operator $\mathcal{L}u$ at the node $\vec{x}_c\in\mathbb{R}^d$ as a linear combination of the function values $u$ at the $n$ closest nodes (known as a stencil),
\begin{equation}
\left.\mathcal{L}u\right|_{\vec{x}_c} \approx \sum_{i=1}^n w_i u_i.
\label{rbffd}
\end{equation}
The main difference lies in how the differentiation weights $w_i$ are computed. While FD enforces \eqref{rbffd} to be exact for polynomials evaluated at the node $\vec{x}_k$, RBF-FD enforces it for RBF interpolants
\begin{equation}
s(\vec{x})=\sum_{i=1}^n\lambda_i\phi\left(\left\|\vec{x}-\vec{x}_i\right\|_2\right),
\label{RBF_interp}
\end{equation}
where $\phi$ is a radial basis function, $\left\|\cdot\right\|_2$ is the Euclidean distance and $\lambda_i$ are the RBF coefficients. Unlike FD, in which the interpolation problem is not guaranteed to be non-singular for scattered nodes in dimensions greater than one, RBF-FD is guaranteed to be non-singular no matter how the $n$ nodes (assumed distinct) are scattered in any number of dimensions \cite{FFbook2015,Fasshauer2007}. 

In developing this novel numerical solver we use 3D PolyHarmonic Spline RBFs (PHS) of order 5, i.e. $\phi=\left\|\mathbf{x}_1-\mathbf{x}_1\right\|_2^5$, augmented with polynomials up to degree 4. For details on using polynomials with RBFs, see Sections 3.1.3.5 in \cite{FFbook2015}. As shown in \cite{FlyerGreg16,Flyer16} and proven in \cite{bayona19}, a smooth target function is approximated only by the order polynomials used. The PHS RBFs only serve to guarantee non-singularity for scattered nodes. As an example, let us calculate the differentiation weights (only first derivatives are needed to form the MHS equations), for approximating $L=\frac{\partial}{\partial y}$ with PHS and polynomials up to degree $1$ (in 3D, note there are 35 polynomials up to degree 4 as we are using in this study). This leads to the following linear system 
\begin{align}
\left[\begin{array}{ccc|cccc}
\left\|\mathbf{x}_1-\mathbf{x}_1\right\|_2^5 & \cdots & \left\|\mathbf{x}_1-\mathbf{x}_n\right\|_2^5 & 1      & x_1    & y_1  & z_1  \\
\vdots                                       & \ddots & \vdots                                       & \vdots & \vdots & \vdots \\
\left\|\mathbf{x}_n-\mathbf{x}_1\right\|_2^5 & \cdots & \left\|\mathbf{x}_n-\mathbf{x}_n\right\|_2^5 & 1      & x_n    & y_n  & z_n  \\
\hline
1                                            & \cdots & 1                                            & 0      & 0      & 0   & 0   \\
x_1                                          & \cdots & x_n                                          & 0      & 0      & 0   & 0   \\
y_1                                          & \cdots & y_n                                          & 0      & 0      & 0   & 0    \\
z_1                                          & \cdots & z_n                                          & 0      & 0      & 0   & 0    \\
\end{array}\right]
\left[\begin{array}{c}
w_1     \\
\vdots  \\
w_n     \\
\hline
w_{n+1} \\
w_{n+2} \\
w_{n+3}  \\
w_{n+4}
\end{array}\right]
=&
\left[\begin{array}{c}
\left.L\left\|\mathbf{x}-\mathbf{x}_1\right\|_2^5\right|_{\mathbf{x}=\mathbf{x}_c}    \\
\vdots                                                                                \\
\left.L\left\|\mathbf{x}-\mathbf{x}_n\right\|_2^5\right|_{\mathbf{x}=\mathbf{x}_c}    \\
\hline
\left.L1\right|_{\mathbf{x}=\mathbf{x}_c} = 0                                          \\
\left.Lx\right|_{\mathbf{x}=\mathbf{x}_c} = 0                                          \\
\left.Ly\right|_{\mathbf{x}=\mathbf{x}_c} = 1                                          \\
\left.Lz\right|_{\mathbf{x}=\mathbf{x}_c} = 0                                          
\end{array}\right].
\label{weights}
\end{align}
The weights $w_{n+1}$ to $w_{n+4}$ are discarded after the matrix is inverted. Solving (\ref{weights}) will give one row of the differentiation matrix $D_y$ that contains the weights for approximating $L$ at $x_c$. Since the stencil size is much less than the total number of discretization nodes (here, by at least three orders of magnitude), $D_y$ is over $99\%$ zeros. As a result, we do not actually store the $D_y$ but only its nonzero entries. The above also holds when calculating $D_x$ and $D_z$. For further in depth details on the calculation of RBF-FD weights, see Section 5.1.4 in \cite{FFbook2015}.

\section{Stencil Setup for MHS}

\begin{figure}
    \centering
    \includegraphics[width=0.45\textwidth]{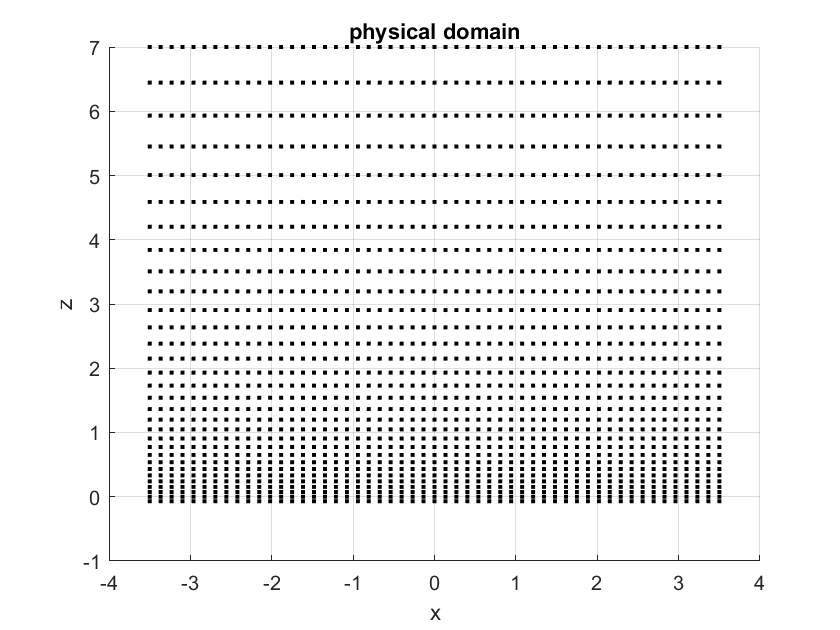}
    \includegraphics[width=0.45\textwidth]{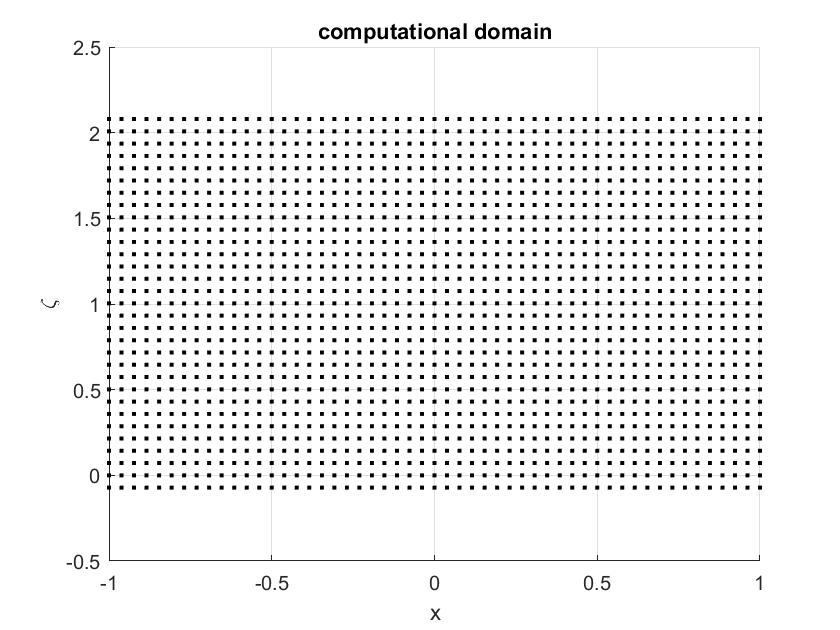}
    \caption{Left: Node layout in the exponentially stretched physical domain as seen in the $z x$ plane. Right: Node layout in the computational domain as seen in the $\zeta x$ plane with $\Delta \zeta \approx \Delta x$. For display purposes, the number of nodes shown is $N\approx 30^3$ with the final computations having close to an order of magnitude more.}\label{nodeset}
\end{figure}

In Cartesian coordinates $\{x,y,z\}$ using array notation, the MHS equations given in (\ref{eq_mhs}) are
\begin{align}\label{eq_mhsvec}
\left(\begin{array}{c c c}
0 & -\partial_z & \partial_y \\ \partial_z & 0 & -\partial_x \\ -\partial_y & \partial_x & 0
\end{array}\right) \left(\begin{array}{c}B_x \\ B_y \\ B_z\end{array}\right)
\times 
\left(\begin{array}{c} B_x \\ B_y \\ B_z\end{array}\right) = \left(\begin{array}{c}F_x\\F_y\\F_z\end{array}\right).\\
\left(\begin{array}{c }
\partial_x  \\ \partial_y \\ \partial_z
\end{array}\right) \mathbf{\cdot} \left(\begin{array}{c}B_x \\ B_y \\ B_z\end{array}\right) =  0.
\end{align}
Expanding the equations results in the following set of static but non-elliptic, first-order PDEs
\begin{align}
\begin{array}{ccccc}
  B_z(\partial_zB_x  - \partial_xB_z) &-& B_y(-\partial_yB_x + \partial_xB_y) &=& F_x\\
 -B_z(-\partial_zB_y + \partial_yB_z) &+& B_x(-\partial_yB_x + \partial_xB_y) &=& F_y\\
  B_y(-\partial_zB_y + \partial_yB_z) &-& B_x(\partial_zB_x  - \partial_xB_z) &=& F_z\\
 &&\partial_xB_x + \partial_yB_y + \partial_zB_z &=& 0. 
\end{array}
\label{eq_cartmhs2}
\end{align}
Since most of the magnetic structure is localized near the lower boundary where much of the information originates, we perform a vertical exponential stretching (i.e. a change of variables), $z =e^{\omega \zeta}-1$, to concentrate nodes near the lower boundary in the physical $z$ direction while they remain equi-spaced in the computational $\zeta$ direction, as shown in Figure \ref{nodeset}. In other words, $D_z = \omega e^{\omega \zeta}D_{\zeta}$. Values of $\omega$ between 0 and 3 were tested with 2 yielding the ideal results for this case study.

\begin{figure}
    \centering
    \includegraphics[clip, trim=5.5cm 9cm 5cm 8cm,width=0.5\textwidth]{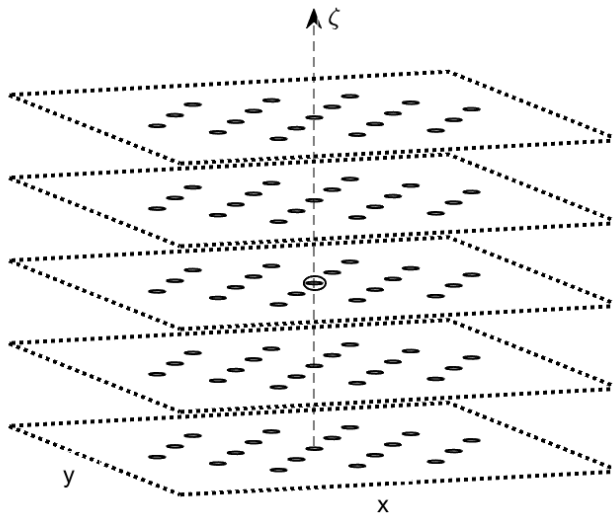}
    \caption{A typical 3D interior stencil, using a hexagonal layout in $x$ and $y$, for approximating $L$ at $x_c$, the circled node. It is composed of 5 planes in the $\zeta$ plane, each plane with a 19 node layout for a total of $N = 95$ nodes. At the boundaries the stencil would become one sided with 31 nodes per plane for a stencil size of 155.}
\label{3Dstencil}
\end{figure}

It was shown in \cite{FlyerGreg16} that hexagonal node layouts in 2D performed better than Cartesian or quasi-uniform layouts when capturing complex structures. We take our cue from that paper, constructing 3D stencils from stacked $xy$-planes of hexagonal nodes, where $\sqrt{3}/2\Delta x \approx \Delta y$ and  $\Delta x \approx \Delta \zeta$. At each interior node, the $19$ nearest neighbors (or $18$ not including itself), are chosen to create a symmetric, hexagonal shape. Next, we stack five such hexagonal planes on top of each other, with regular spacing in $\zeta$. This forms a $19\times 5 = 95$ 3D stencil at which the RBF weights for a derivative operator are calculated at the center node, as shown in Figure \ref{3Dstencil}. For boundary stencils, we do the same procedure except we use the $31$ nearest neighbors to create a $31\times 5 = 155$ 3D stencil. The rationale is that we can avoid accuracy deterioration near the boundaries by making the stencils larger as was shown in \cite{BAYONAIII}. Here, the advantage is that we only need to calculate the weights for one stencil, be it a boundary or interior stencil, and shift it by the node spacing to get the entire differentiation matrix, making pre-processing costs trivial.

\section{A General Numerical Algorithm for Solving MHS}\label{section_numimp}

\subsection{Overview}
We solve the system defined in (\ref{eq_cartmhs2}) with a Quasi-Newton approach. This sort of solver begins with an initial guess, $\mathbf{B}_0$, for the magnetic field, and then updates it by calculating its distance from MHS balance, or \emph{residual}. We use an iterative solver, LSQR \cite{lsqr}, for sparse tall overdetermined systems to find a solution to the Jacobian system for every update of the residual. The process is then re-iterated until convergence to an acceptable tolerance is achieved. Two points should be noted with regard to the implemented approach: 1) a customized preconditioner is used to accelerate convergence and 2) a divergence ``cleaning'' algorithm is used after the Quasi-Newton method has converged to remove leftover divergence. The initial guess, $\mathbf{B}_0$ needs to be as informed as possible due to the nature of these equations as described in the Introduction. When seeking sufficiently smooth solutions without strong plasma forcing (i.e. right hand side term \textbf{F} is small), the potential field model, $\nabla\times\mathbf{B}=0$ suffices. However, for fields with expected discontinuities, as will be the case in our numerical study, it is useful to incorporate the discontinuity into the initial guess $\mathbf{B}_0$. The below steps overview the numerical process.

\begin{enumerate}
    \item Given $\mathbf{B}_0$, construct a residual vector field $\mathbf{r}$ by inserting the initial guess into the MHS  PDEs, which measure the degree to which $\mathbf{B}_0$ deviates from MHS balance.
    \item Construct the Jacobian matrix $\mathcal{J}$ which is known analytically for a given set of differentiation matrices. In our case $\mathcal{J}$ is composed of 3D PHS RBFs plus polynomials.
    \item Supplement the matrix $\mathcal{J}$, which otherwise has a nontrivial nullspace, with the divergence-free constraint to obtain an overconditioned tall system.
    \item Use the LSQR iterative solver that calls a custom preconditioner to invert the Jacobian update, obtaining an update vector $\mathbf{u}$.
    \item Update the magnetic field according to $\mathbf{B}_{\text{new}} = \mathbf{B}_{\text{old}}+\mathbf{u}$. Repeat steps 1-5 until convergence to a desired level is achieved.
    \item Remove any remaining divergence in the magnetic field without perturbing the solution from the Quasi-Newton solver. This is done by assuming a conservative field with identical divergence to whatever divergence is present in the numerically constructed field, i.e. $\nabla\cdot\mathbf{B}_N=\nabla^2\phi$, leading to a Poisson type equation that is solved via General Mean Residual (GMRES). A similar situation occurs when solving for pressure in incompressible fluid flows, ensuring that continuity is satisfied. 
\end{enumerate}

\subsection{Construction of The Residual and Jacobian}

We wish to quantify the pointwise distance from MHS balance for a given magnetic field. To that end, we construct the vector-valued Residual function via the RBF-FD discretization. Let $D_e$ be the differentiation matrix in direction $\hat{e}$. 
For a given magnetic field $\mathbf{B} = \langle B_x,\,B_y,\,B_z\rangle$, define the curl of $\mathbf{B}$ (also known as the current and hence we use $\mathbf{J}=\langle J_x,\,J_y,\,J_z\rangle$ as its symbol) as
\begin{equation}
    \mathbf{J} = \left(\begin{array}{c}J_x\\J_y\\J_z\end{array}\right)
        = \frac{1}{4\pi}\left(\begin{array}{c c c}
        0 & -\omega e^{\omega \zeta}D_{\zeta} & D_y\\
        \omega e^{\omega \zeta}D_{\zeta} & 0 & -D_x\\
        -D_y & D_x & 0
        \end{array}\right)\left(\begin{array}{c}B_z\\B_y\\B_z\end{array}\right).
\end{equation}
We then construct the residual vector as
\begin{equation}\label{eq_res}
\begin{split}
        \mathbf{r} &= \left(\begin{array}{c}r_x\\r_y\\r_z\end{array}\right)
        = \left(\begin{array}{c c c}
        0 & -J_z^d & J_y^d\\
        J_z^d & 0 & -J_x^d\\
        -J_y^d & J_x^d & 0
        \end{array}\right) \left(\begin{array}{c}B_z\\B_y\\B_z\end{array}\right) - 
        \left(\begin{array}{c}F_x\\F_y\\F_z\end{array}\right) - \gamma\mathbf{H_V}\\
\end{split}
\end{equation}
where the superscript $d$ denotes a diagonal matrix formed from the corresponding component of $\mathbf{J}$. $\mathbf{H_V}$ is an introduced $3\text{x}3$ block diagonal hyperviscosity matrix. The entries along the diagonal are the 3D Laplacian in $\{x,y,\zeta\}$ given by $D_{xx} + D_{yy} + \omega e^{2 \omega \zeta} D_{\zeta\zeta} - \omega e^{\omega \zeta} D_\zeta$. Furthermore, it is multiplied by a parameter $\gamma$, controlling how much hyperviscosity is added. Experimental results indicate $\gamma$ of $O(10^{-4})$ is suitable. The analytic Jacobian $\mathcal{J}$ is calculated by taking the analytic derivative of the numerical discretization given in \eqref{eq_res} with regard to each unknown. For compactness of notation, let $\alpha = \omega e^{\omega \zeta}$, the term that occurs from doing the change of variables from $\frac{\partial }{\partial z}$ to $\frac{\partial }{\partial \zeta}$. The Jacobian is then defined as 
\begin{align}\label{eq_jacobian}
    \begin{array}{l l l}
       \mathcal{J} =  \frac{1}{4\pi}\left(\begin{array}{l l l}
        B_y^dD_y + B_z^d\alpha D_{\zeta} & (D_yB_x-D_xB_y)^d-B_y^dD_x & \alpha D_{\zeta}B_x-D_xB_z)^d-B_z^dD_x\\
        (D_xB_y-D_yB_x)^d-B_x^dD_y & B_x^dD_x + B_z^d\alpha D_{\zeta} & \alpha D_{\zeta}B_y-D_yB_z)^d-B_z^dD_y\\
        (D_xB_z-\alpha D_{\zeta}B_x)^d-B_x^dD_z & (D_yB_z-\alpha D_{\zeta}B_y)^d-B_y^d\alpha D_{\zeta} & B_x^dD_x + B_y^dD_y
    \end{array}\right) \\
    \end{array}
\end{align}

As with the residual in (\ref{eq_res}), we also add a hyperviscosity term $-\gamma\mathbf{H_V}$ to (\ref{eq_jacobian}). It should be noted that using  hyperviscosity in an elliptic type solver is not a common procedure. However for ill-conditioned problems as in magnetostatics or steady flows, where discontinuities appear or the solution space fundamentally changes in the limit as a parameter/forcing goes to zero, hyperviscosity greatly helps in adding numerical stability.

\subsection{Solving The Discretized System}

The Jacobian in (\ref{eq_jacobian}) is generally not full rank. To find an update vector $\mathbf{u}_n=\langle u_x, u_y, u_z\rangle$ at the $n^{\text{th}}$ iteration of the Quasi-Newton method such that $\mathcal{J}_n\mathbf{u}_n = -\mathbf{r}_n$, it is necessary to eliminate the null space of $\mathcal{J}_n$ by including additional constraint rows. We choose the divergence-free property as the constraint since it is imposed by the physics of the equations. This leads to the following over-determined system
\begin{equation}\label{eq_update}
\left(\begin{array}{c c c} & \mathcal{J}_n\\
\varepsilon D_x & \varepsilon D_y & \varepsilon \omega e^{\omega \zeta}D_{\zeta}
\end{array}\right) \left(\begin{array}{c}
u_x\\u_y\\u_z
\end{array}\right) = 
-\left(\begin{array}{c}\mathbf{r_n}\\
\varepsilon (D_xB_x + D_yB_y + \omega e^{\omega \zeta}D_{\zeta}B_z)
\end{array}\right) 
\end{equation}
where $\varepsilon>0$ (note, $\varepsilon$ can not be factored out as the system is solved in a least-squares sense). When $\varepsilon$ is too large, it is no longer ensured that $\mathbf{u}_n$ will move the field towards MHS balance. However, if $\varepsilon$ is too small, (\ref{eq_phibound}) could be quite large, and guarantees on the perturbation introduced by the divergence cleaning step maybe lost. In numerical studies, $\varepsilon=0.1$ was found to have a good balance between these factors.

\begin{figure}
    \centering
    \includegraphics[trim = 50 0 50 0, clip, width=0.3\textwidth]{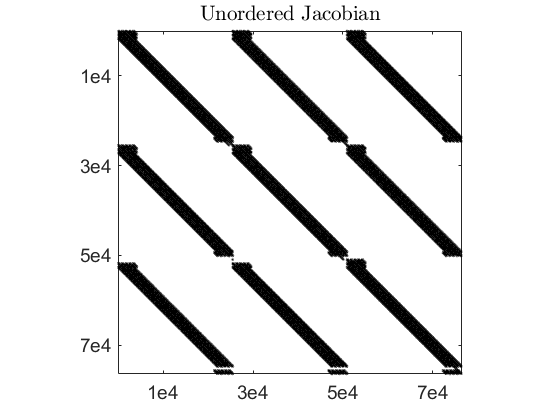}
    \includegraphics[trim = 50 0 50 0, clip, width=0.3\textwidth]{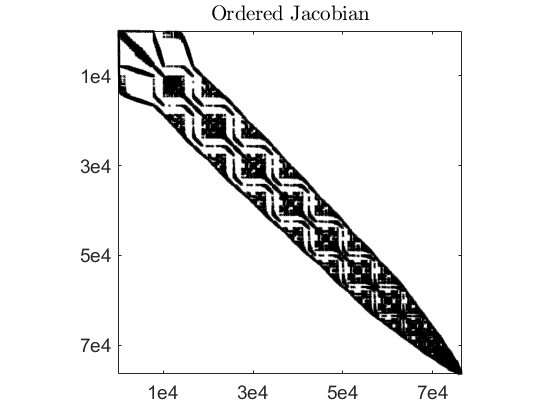}
    \includegraphics[trim = 50 0 50 0, clip, width=0.3\textwidth]{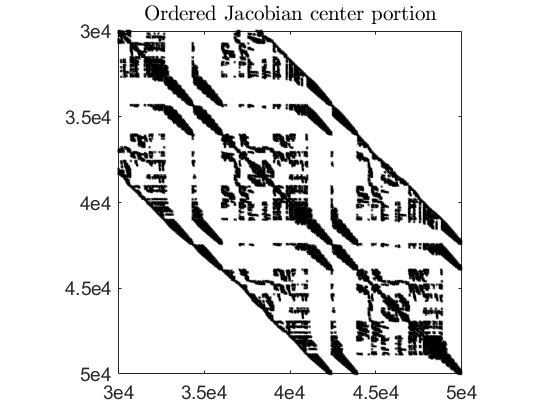}
    \caption{The sparsity structure of the RBF-FD Jacobian with approximately $30^3$ computational points. On the left is the Jacobian organized with the native node ordering (the nodes were ordered from the top of the domain downwards, thus the banding). The middle figure is the same Jacobian but reordered via a symmetric reverse Cuthill-McKee algorithm. On the right is a zoom-in to investigate the substructure of the ordered Jacobian. The matrix is $0.4\%$ nonzeros.}
    \label{fig_jacobian}
\end{figure}

Applying a least-squares method to RBF-FD has been used on the local stencil level to increase numerical stability, especially for Neumann boundary conditions \cite{tominec2021squares}. Here, a least squared approach is used in terms of applying an LSQR iterative solver to invert the tall (i.e. overdetermined) Jacobian matrix. In order to increase the speed of the method,  a symmetric reverse Cuthill-McKee reordering on $\mathcal{J}$ is performed. Figure \ref{fig_jacobian} demonstrates how the $\mathcal{J}$ goes from a block banded structure throughout the entire matrix to a centrally-banded diagonal structure.  Also, a preconditioner is applied to the system. With regard to the latter, the common incomplete $LU$ factorization is not a good choice here since it becomes highly memory intensive and ineffective as the resolution increases. Instead,  we precondition the system by an approximation of the background magnetic field. For such an approximation, the potential field satisfying
\begin{equation}
    \nabla\times\mathbf{B}=0,\,\nabla\cdot\mathbf{B}=0.
    \label{eq_pot}
\end{equation}
is selected. Note that a potential field does not have to be a linear field as normally thought of, but can have discontinuities as in our study and still satisfy (\ref{eq_pot}). 

\subsection{Divergence Cleaning} 
\label{sec_divclean}
After the Quasi-Newton method has converged to a final magnetic field $\mathbf{B}_N$, we clean away remaining divergence that might be present in the field by constructing a conservative field with the same divergence, and subtracting it away. This may perturb $\mathbf{B}_N$ slightly away from force-balance, but we justify this by assuming that MHS is an \emph{approximation} to a truly magnetohydrodynamic system on long time scales.

This assumption leads to the following Poisson type equation for conservative field $\nabla\phi$:
\begin{equation}
    \nabla^2\phi = \nabla\cdot\mathbf{B}_N.
    \label{eq_poisson}
\end{equation}
\eqref{eq_poisson} is solved by a GMRES algorithm. For this purpose, we take the boundary conditions
\begin{equation}
\begin{split}
    \phi(z=z_{max}) &= 0\\
    D_z\phi(z=0) &= 0\\
    \text{$\phi$ is periodic} & \text{ in $x$ and $y$ }  
\end{split}
\end{equation}
To bound the amount this cleaning process perturbs the field $\mathbf{B}_N$ from MHS balance, we plug the cleaned field $\mathbf{B}_N-\nabla\phi$ into the MHS balance equation:
\begin{equation}\label{eq_phibound}
\left(\nabla\times(\mathbf{B}_N-\nabla\phi)\right)\times(\mathbf{B}_N-\nabla\phi) = (\nabla\times\mathbf{B}_N)\times\mathbf{B}_N-(\nabla\times\mathbf{B}_N)\times(\nabla\phi) = \mathbf{F}-(\nabla\times\mathbf{B}_N)\times(\nabla\phi).
\end{equation}
By virtue of the selection of boundary conditions on $\phi$, we can ensure that $||\nabla\phi||_2$ is bounded for smooth problems by $L\nabla\cdot\mathbf{B}_N$, for some constant $L$. This bounds the total perturbation by $L(\nabla\cdot\mathbf{B}_N)\,||\nabla\times\mathbf{B}||$. This is a small perturbation because $\nabla\cdot\mathbf{B}_N$ is small, having been included in the overconditioned least-squares updates. 

\section{Numerical Studies on Highly Nonlinear MHS Fields}\label{section_Testbed}

\begin{figure}[h!]
    \centering
    \includegraphics[width=0.35\textwidth]{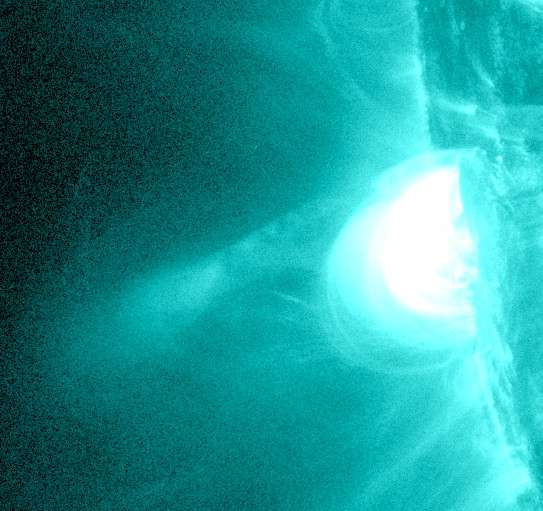}
    \caption{An observation presented by \textcite{nindos20} as of a flux rope, as seen in 131 {\AA} from the Solar Dynamics Observatory (SDO) satellite on 2014-02-24 at 11:45:58 UT. This and other data from the SDO instrument can be found at \url{https://sdo.gsfc.nasa.gov/data/aiahmi/}.}
    \label{fig_obs}
\end{figure}

We demonstrate our solver by reconstructing the analytic static solution from \textcite{gibson98}, overviewed in Apppendix A. This model is meant to mimic the complex and twisted structure of flux ropes in the solar corona. Though the original model paper includes a dynamic system, an equilibrium solution is detailed in Appendix 2 thereof. This MHS solution has very tightly bundled magnetic fields with strong, nonlinear currents which can be difficult to model. The strong forces in this simulation are exactly why it is frequently used in literature, as for example an initial state for dynamic simulations of coronal mass ejections in the context of space weather forecasting \cite{dove11,gibson00,lugaz13,malanushenko20,manchester03,Manchester04}. These toroidal, or spheromak, flux ropes can form as a result of the kink instability \cite{torok05} and potentially exist in the corona either as a quasistable magnetic structure \cite{dove11} or in eruption \cite{gibson08}. Observational evidence for magnetic flux ropes have been found before, during and after eruptions, e.g. Figure~\ref{fig_obs}. Compare this observation to the analytic solution from \textcite{gibson98} given in Figure~\ref{fig_giblowSide}~e.

We will consider two different emergences of the Gibson-Low model -- one where the flux rope lies near the solar surface, and another where it is higher up, forming a torus whose central axis is fully out of the plane. See \textcite{malanushenko20} for a discussion of how emergence results in changes in magnetic topology. The vector-valued function that forces the more emerged case is shown in Figure~\ref{fig_giblowPlasma} and depends upon the density and pressure defined in Appendix A by \eqref{eq_GLsoln}. The model is highly nonlinear in both cases, making it a good test for the solver. We provide the solver with full information about the plasma forcing, as well as a Dirichlet vector-valued lower boundary condition, radial-field side boundaries and a Neumann upper boundary. In particular:
\begin{equation}
\begin{split}
    B_x(z=0) &= B_{x0}\\
    B_y(z=0) &= B_{y0}\\
    B_z(z=0) &= B_{z0}\\
    B_x(x=-1)^2+B_y(x=-1)^2 &= 0\\
    B_x(x=1)^2+B_y(x=1)^2 &= 0\\
    B_x(y=-1)^2+B_y(y=-1)^2 &= 0\\
    B_x(y=1)^2+B_y(y=1)^2 &= 0\\
    D_z B_z(z=z_{max}) &= 0
\end{split}
\end{equation}
where $B_{x0}$, $B_{y0}$ and $B_{z0}$ are defined by $\mathbf{B}(z=0)$ in the analytic model (we are assuming that we know the full magnetic vector field at the $z=0$ boundary), which has discontinuities.

\begin{figure}[h!]
    \includegraphics[width=0.33\textwidth]{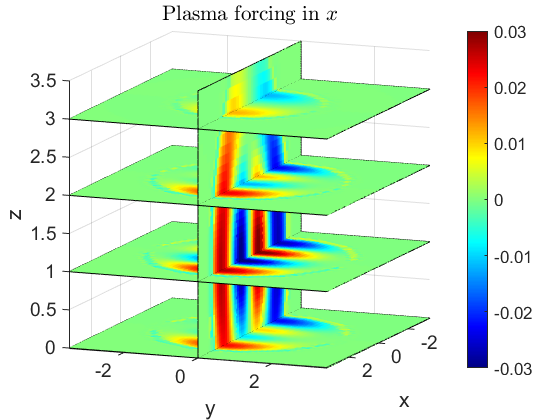}
    \includegraphics[width=0.33\textwidth]{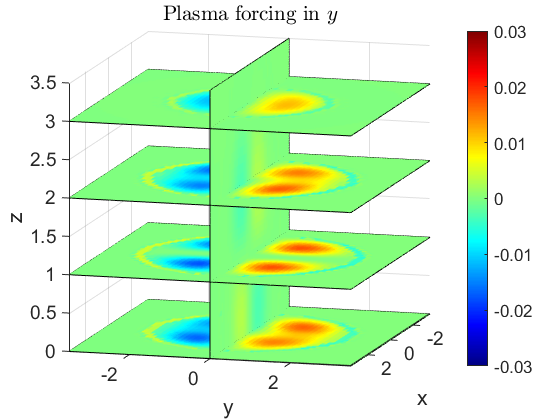}
    \includegraphics[width=0.33\textwidth]{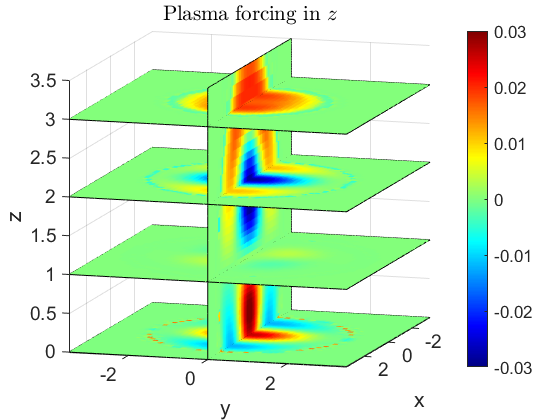}
    \caption{The vector-valued plasma forcing function $\mathbf{F}=\nabla P + \rho g \hat{z}$ that drives the Gibson-Low model in the $x,y,z$ directions, correspondingly, from left to right. $P$ and $\rho$ are given in Appendix~\ref{Appendix_GibLow}.}
    \label{fig_giblowPlasma}
\end{figure}

\subsection{Comparison of Numerical Results to Analytic Solution}

\begin{figure}
\centering
    \includegraphics[width=0.45\textwidth]{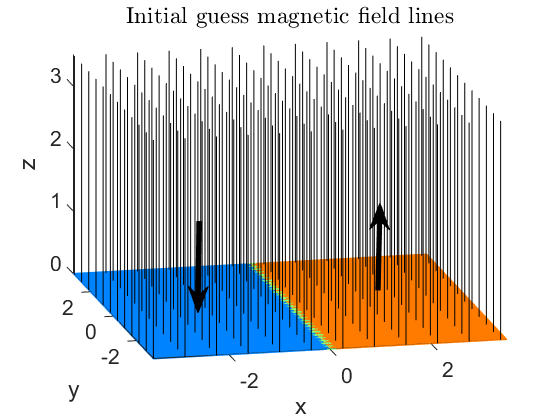}
    \caption{The initial guess used for the $N\approx 30^3$ numerical simulation. Blue represents a $B_z$ value of $-0.1$ and orange a value of $0.1$.}\label{fig_initguess}
\end{figure}

 The delicate nature of solving such a highly nonlinear MHS field requires increments in resolution when applying the initial guess, yet incorporating also the basic structure of the magnetic field. For a low resolution model, $N=25482\approx30^3$, we take as an initial guess a field that matches the background magnetic field in the Gibson and Low model. This is a purely vertical field with $B_z=0.1$ in the positive-$x$ half-plane (orange) and $B_z=-0.1$ in the negative-$x$ half-plane (blue), having an inherent discontinuity at $x=0$; see Figure \ref{fig_initguess}. The numerical solution at this low resolution is then itself interpolated onto a finer $N=195383\approx60^3$ node set and used as the initial guess to re-run the solver at that resolution.\footnote{The leading order error of a first derivative approximation has a dissipative character as $O(h^2), O(h^4)$, etc, where $h$ is the grid spacing. As the grid is refined these terms that may have a controlling effect fade out.} In both runs, the preconditioning field for the Quasi-Newton Jacobian system is the vertical background field. The Quasi-Newton method is capped at twenty iterations but for all resolutions the convergence criterion is reached before that threshold. 

\begin{figure}[h!]
    \centering
    \includegraphics[width=0.45\textwidth]{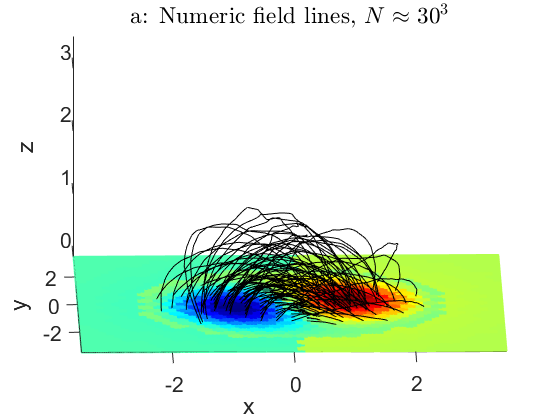}
    \includegraphics[width=0.45\textwidth]{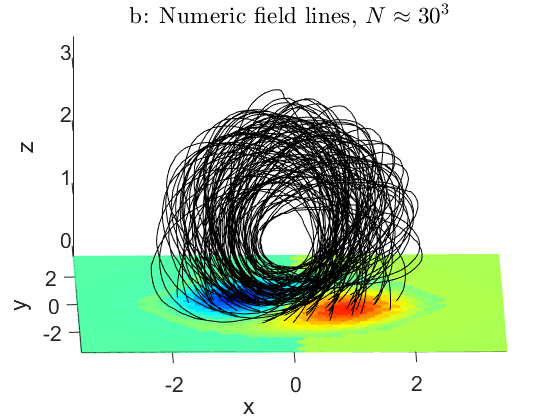}\\
    \includegraphics[width=0.45\textwidth]{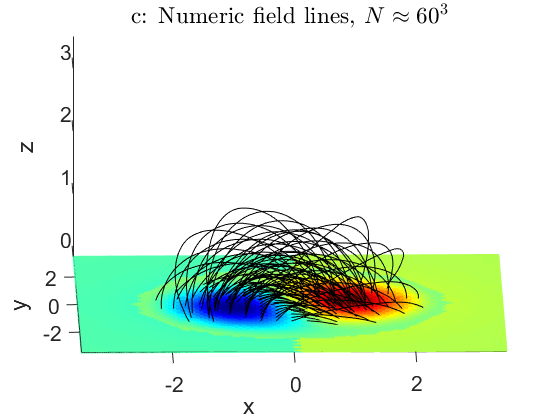}
    \includegraphics[width=0.45\textwidth]{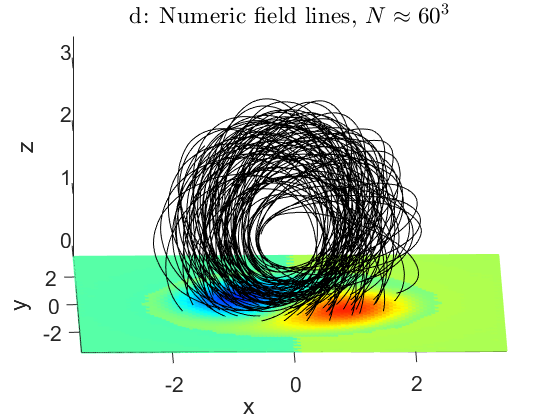}\\
    \includegraphics[width=0.45\textwidth]{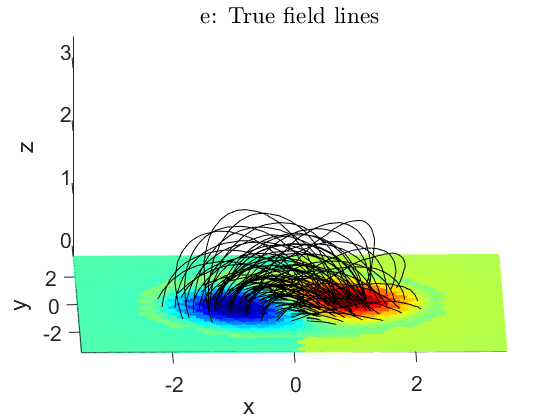}
    \includegraphics[width=0.45\textwidth]{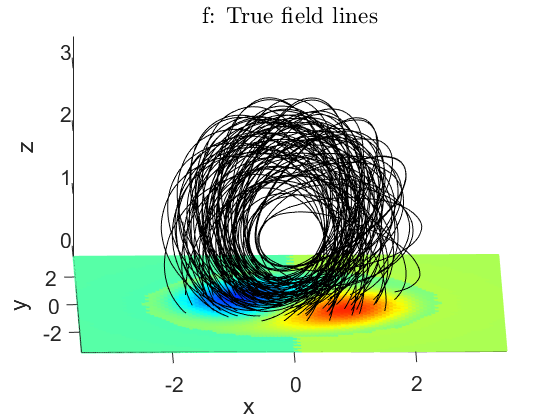}
    \caption{A side view of the magnetic field lines in the Gibson-Low structure for two different emergence levels (left and right columns). We omit the field lines that originate outside a circle centered at the origin in order to isolate the structure for plotting purposes. On the top row is shown the results for the method with $N\approx 30^3$ points, which used vertical up-and-down field lines as the initial guess. The second row is $N\approx60^3$ points, which used an interpolation of the $30^3$ case above as its initial guess.}\label{fig_giblowSide}
\end{figure}

\begin{figure}[h!]
    \centering
    \includegraphics[width=0.45\textwidth]{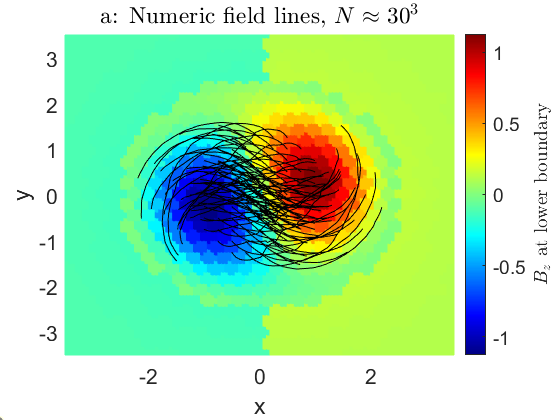}
    \includegraphics[width=0.45\textwidth]{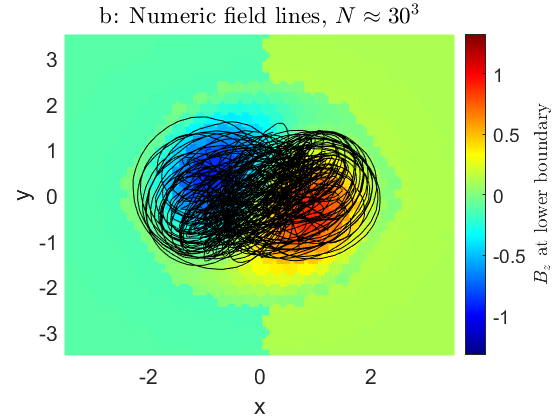}\\
    \includegraphics[width=0.45\textwidth]{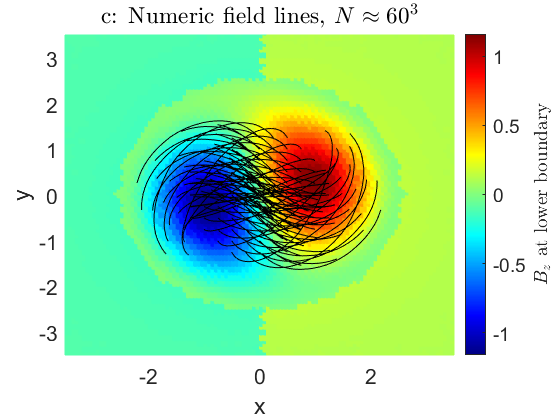}
    \includegraphics[width=0.45\textwidth]{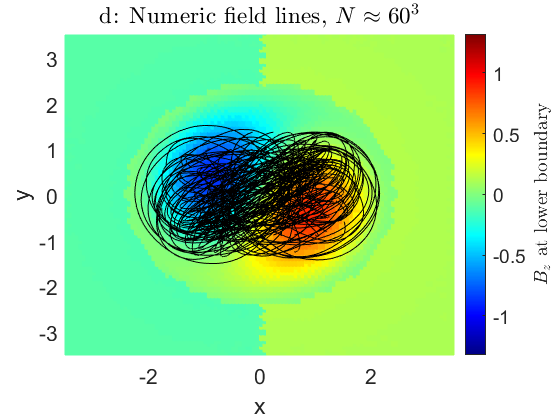}\\
    \includegraphics[width=0.45\textwidth]{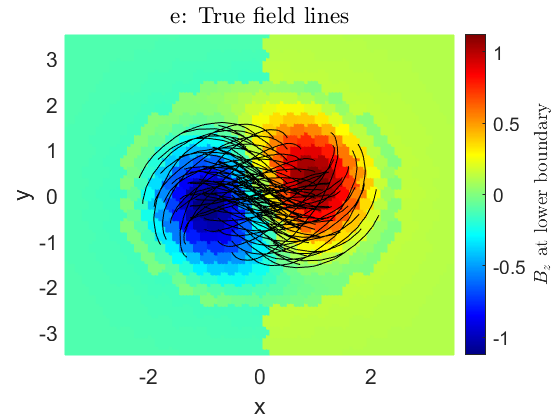}
    \includegraphics[width=0.45\textwidth]{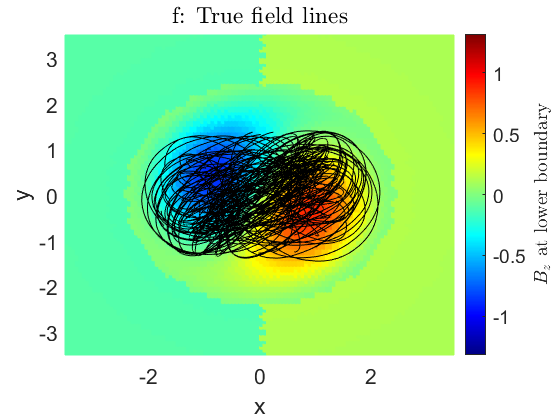}
    \caption{Shown here are top views of the same fields displayed in Figure \ref{fig_giblowSide}, in the same configuration.}\label{fig_giblowTop}
\end{figure}

 Figures \ref{fig_giblowSide} and \ref{fig_giblowTop} display the solver results and the true solution (e and f) from a side and top view, respectively. With regard to the relatively sparse $N\approx 30^3$ node set used, the full structure of the field is generally recovered, but the field lines are not quite as tightly bundled as the true solution. The field lines appear to be slightly kinked and rough, especially as they approach the boundary of the closed structure where a discontinuity exists. This is true for both the simple flux rope (left-hand images) and the more complicated emerged-spheromak configuration (right-hand images). With regard to the higher resolution test case of $N\approx 60^3$ nodes set, it is clear that the refinement has been extremely effective. All of the ways in which the coarse simulation differed from the analytic solution have been addressed, with the numerical solution being an excellent reproduction. The only remaining differences are slight deviations in footpoint locations where the field lines return to the lower boundary.
 
\subsection{Performance of the MHS Solver in the Context the Gibson and Low Model}

In this section, we discuss the details of the general algorithm outlined in Section~\ref{section_numimp} as applied to the Gibson and Low model. We will analyze the effects of preconditioning on convergence of the LSQR iterative solver and overall Quasi-Newton method, divergence removal, and computational time.

\begin{figure}
    \centering
    \includegraphics[width=0.45\textwidth]{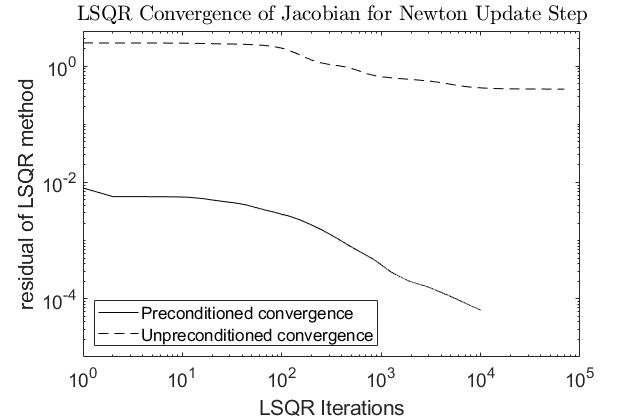}
    \caption{The convergence of LSQR, with and without preconditioning, for the tall system in \eqref{eq_update} at an arbitrarily chosen Quasi-Newton step for $N\approx60^3$.}
    \label{fig_lsqr}
\end{figure}

Figure \ref{fig_lsqr} shows the convergence of the LSQR method for the $N\approx 60^3$ case, with and without the preconditioner - the vertical background field of Figure \ref{fig_initguess}, in terms of the residual at an arbitrary Quasi-Newton step. Without preconditioning, no convergence is achieved. With preconditioning the convergence is approximately linear with $10^4$ iterations used at each step. The number of LSQR iterations for convergence of a sparse matrix is proportional to the matrix size.

\begin{figure}
    \centering
    \includegraphics[width=0.45\textwidth]{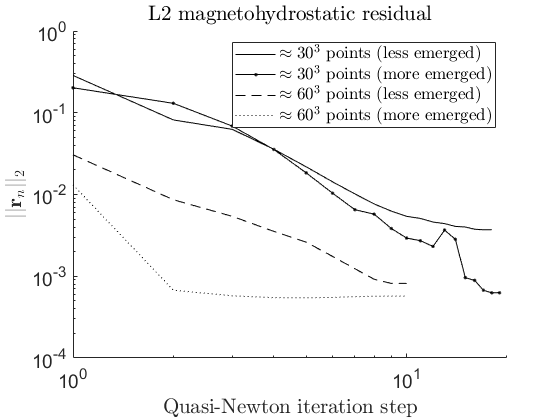}
    \includegraphics[width=0.45\textwidth]{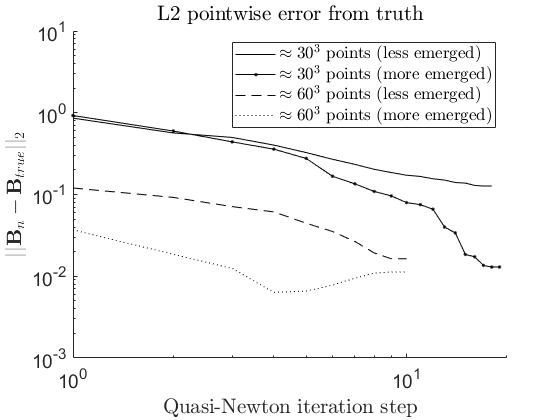}
    \caption{Left: Convergence of the Quasi-Newton's method, in terms of the MHS residual (\ref{eq_res}). The norm squared is summed over each nodepoint inside the flux rope (where the forcing is nonzero, see Figure~\ref{fig_giblowPlasma}), divided by the number of such points, and then the square root is taken. Right: The same measurement technique is applied to the vector magnetic field error.}\label{fig_convergence}
\end{figure}

Figure~\ref{fig_convergence} shows the convergence of the Quasi-Newton solver as measured by the $L2$ norm of two quantities, sampled at each location inside the magnetic flux rope. The flux rope boundaries for this purpose were determined by considering the region of nonzero $\mathbf{F}$. On the left, the MHS residual is considered. Since the nodes are clustered near the lower boundary in the physical domain, this metric weights error more heavily towards the lower boundary where most of the magnetic structure is localized. Convergence is roughly second order in the $N\approx30^3$ case -- the residual shrinks by approximately two orders of magnitude in $10$ iterations, and reaches a convergence criterion in fewer than $20$ iterations. As can be seen in this figure, the higher-resolution case starts at  a much smaller residual and needs only 10 iterations to fully converge.

The right-hand panel of the same figure considers the error of the vector field as of a given Quasi-Newton step when compared to the true magnetic field, which is known for this testbed problem \cite{gibson98}. An error of less than $2\%$ is attained by the end of the routine in all but one case.

\begin{figure}[h!]
    \centering
    \includegraphics[width=0.45\textwidth]{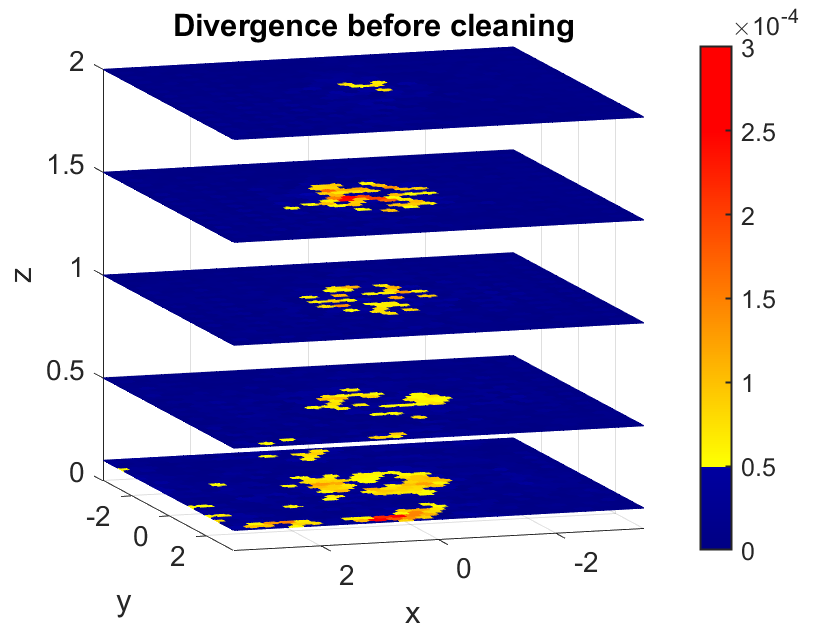}
    \includegraphics[width=0.45\textwidth]{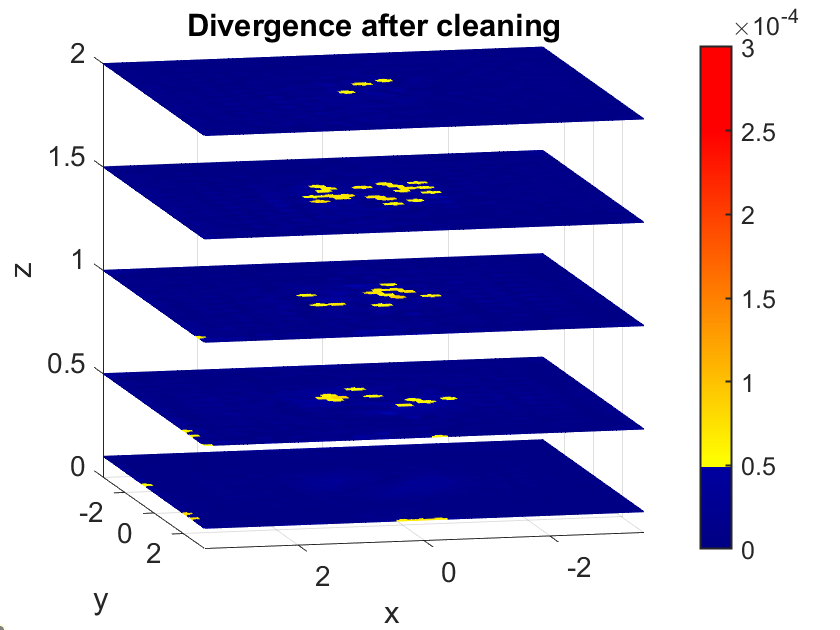}
    \caption{The magnitude of divergence, displayed in the lower part of the physical domain, before and after the Poisson-based cleaning process for the field shown in the left-hand column of Figure~\ref{fig_giblowSide}~b. Above this level, the divergence of MHS solver is O($10^{-10})$, as the field is becomes largely vertical.}
    \label{fig_divremoval}
\end{figure}

After the Quasi-Newton solver has converged, the residual divergence cleaning described in Section \ref{sec_divclean} is applied. In Figure \ref{fig_divremoval}, before applying cleaning the field has a maximum divergence of $O(10^{-4})$ as seen in the left panel of the plot. After applying the cleaning process, the divergence has been reduced by a minimum of one order of magnitude, especially near the lower boundary where as noted before much the magnetic action occurs.

\begin{figure}[h!]
    \centering
    \includegraphics[width=0.45\textwidth]{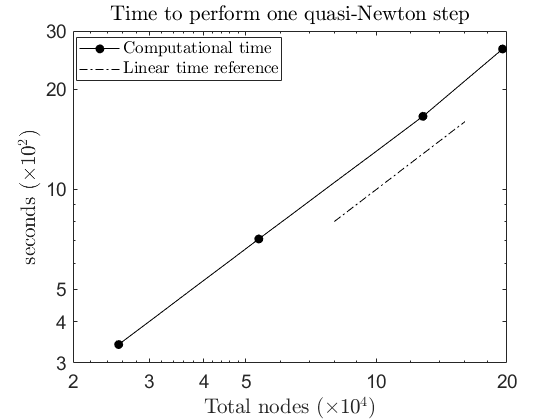}
    \caption{The time to execute a single Quasi-Newton step. We observe an approximately linear relationship between  computational time cost and total grid size.}\label{fig_timing}
\end{figure}

With regard to timing costs, Figure \ref{fig_timing} displays the time required to compute a single Newton update step for a selection of node counts on an AMD Ryzen 7 1800X 3.60GHz processor with 64 GB of RAM. The computation scales linearly with the total number of nodes. The $N\approx 60^3$ case takes about $8$ times longer than the $N\approx 30^3$. It should be noted that the most expensive part of the overall algorithm is the the calculation of a Quasi-Newton update vector by the LSQR solver.

\section{Summary}

We have presented a novel numerical solver, based on 3D RBF-FD PHS plus polynomials, for MHS systems. Due to the nature of the PDEs, as discussed in the bullet points of the Introduction, MHS are a very challenging set of equations to numerically solve. In order to extract a physically viable solution numerically requires:
\begin{enumerate}
    \item Flexibility, simplicity and accuracy in the spatial discretization, as exemplified by RBF-FD PHS plus polynomials
    \item A sparse iterative solver that can be successfully implemented for constrained  hyperbolic-type PDEs, in this case Least-Squares
    \item A fast preconditioner that mimics the background field and can include discontinuities
    \item A residual divergence removal that minimally perturbs the numerically achieved MHS balance.
\end{enumerate}
As a difficult test problem that mimics observations of the sun, we apply our algorithm to the Gibson and Low Model with the following results:
\begin{enumerate}
    \item Being able to reconstruct highly nonlinear magnetic structures in the solar corona
    \item The numerically-constructed magnetic field is divergence free to at least $0.001\%$ of the maximum field strength
    \item The cost, as measured by computational time, linearly scales with the total number of nodes.
\end{enumerate}


\section*{Acknowledgements}
This material is based on work supported by The National Center for Atmospheric Research, which is a major facility sponsored by the National Science Foundation under Cooperative Agreement No. 1852977. This work was also supported in part by the Air Force Office of Scientific Research grant FA9550-15-1-0030.

\printbibliography

@article{zhu18,
    author = {Zhu, X. and Wiegelmann, T.},
    title = {On the Extrapolation of Magnetohydrostatic Equilibria on the Sun},
    journal = {The Astrophysical Journal},
    year = {2018},
    volume = {866},
    pages = {130},
    doi = {10.3847/1538-4357/aadf7f}
}

@article{wiegelmann_neukirch06,
	author = {Wiegelmann, T. and Neukirch, T.},
	title = {An optimization principle for the computation  of MHD equilibria in the solar corona},
	doi = {10.1051/0004-6361:20065281},
	journal = {Astronomy \& Astrophysics},
	year = 2006,
	volume = 457,
	number = 3,
	pages = {1053-1058},
}

@article{MacTaggart13,
	author = {MacTaggart, D. and Elsheikh, A. and McLaughlin, J. A. and Simitev, R. D.},
	title = {Non-symmetric magnetohydrostatic equilibria: a multigrid
          approach},
	doi = {10.1051/0004-6361/201220458},
	journal = {Astronomy \& Astrophysics},
	year = 2013,
	volume = 556,
	pages = {A40},
}

@article{mathews20,
    author = {Mathews, N. H. and Flyer, N. and Gibson, S. E.},
    title = {Reconstructing the Coronal Magnetic Field: The Role of Cross-field Currents in Solution Uniqueness},
    journal = {The Astrophysical Journal},
    year = {2020},
    volume = {898},
    pages = {70},
    doi = {10.3847/1538-4357/ab9dfd}
}

@article{gibson98,
    author = {Gibson, S. E. and Low, B. C.},
    title = {A Time-dependent Three-dimensional Magnetohydrodynamic Model of the Coronal Mass Ejection},
    journal = {The Astrophysical Journal},
    year = {1998},
    volume = {493},
    pages = {460},
    doi = {10.1086/305107}
}

@article{tominec2021squares,
      title={A least squares radial basis function finite difference method with improved stability properties},
      author={Tominec, I. and Larsson, E. and Heryudono, A.},
      year={2021},
      journal = {SIAM Journal on Scientific Computing},
      volume = {43},
      number = {2},
      pages = {A1441–A1471},
      doi={10.1137/20M1320079}
}

@article{Shu2003,
	Author = {Shu, C and Ding, H and Yeo, K.S},
	Doi = {10.1016/S0045-7825(02)00618-7},
	Journal = {Computer Methods in Applied Mechanics and Engineering},
	Month = feb,
	Number = {7-8},
	Pages = {941-954},
	Title = {Local radial basis function-based differential quadrature method and its application to solve two-dimensional incompressible Navier--Stokes equations},
	Volume = {192},
	Year = {2003}
}

@article{Tolstykh2003,
	Author = {Tolstykh, A. I. and Shirobokov, D. A.},
	Doi = {10.1007/s00466-003-0501-9},
	Journal = {Computational Mechanics},
	Month = dec,
	Number = {1},
	Pages = {68--79},
	Title = {On using radial basis functions in a {"}finite difference mode{"} with applications to elasticity problems},
	Volume = {33},
	Year = {2003}
}

@article{Wright2006,
	Author = {Wright, G. and Fornberg, B.},
	doi = {10.1016/j.jcp.2005.05.030},
	Journal = {Journal of Computational Physics},
	Pages = {99-123},
	Title = {Scattered node compact finite difference-type formulas generated from radial basis functions},
	Volume = {212},
	Year = {2006}
}

@book{Fasshauer2007,
	Author = { G. F., Fasshauer}, 
	Publisher = {World Scientific Publishers, Singapore},
	Title = {{Meshfree Approximation Methods with MATLAB}},
	Volume = {6},
 	Year = {2007},
}

@book{FFbook2015,
	Author = {B. Fornberg and N. Flyer},
	Publisher = {SIAM Press, Philadelphia, PA},
	Title = {A Primer on Radial Basis Functions with Applications to the Geosciences},
	Year = {2015}
}

@article{bayona19,
      title={An insight into RBF-FD approximations augmented with polynomials},
      author={V. Bayona},
      year={2019},
      journal = {Computers \& Mathematics with Applications},
      volume = {77},
      number = {9},
      pages = {2337-2353},
      doi = {10.1016/j.camwa.2018.12.029}
}

@article{Flyer16,
      title={On the role of polynomials in RBF-FD approximations: I. Interpolation and accuracy},
      author={N. Flyer and B. Fornberg and V. Bayona and G. Barnett},
      year={2016},
      journal = {Journal of Computational Physics},
      volume = {321},
      pages = {21-38},
      doi = {10.1016/j.jcp.2016.05.026}
}

@article{FlyerGreg16,
      title={Enhancing finite differences with radial basis functions: experiments on the Navier–Stokes equations},
      author={N. Flyer and G. Barnett and L. Wicker},
      year={2016},
      journal = {Journal of Computational Physics},
      volume = {316},
      pages = {39-62},
      doi = {10.1016/j.jcp.2016.02.078}
}

@article{Low_2007,
	doi = {10.1086/520503},
	year = 2007,
	publisher = {American Astronomical Society},
	volume = {668},
	number = {1},
	pages = {557-570},
	author = {B. C. Low and N. Flyer},
	title = {The Topological Nature of Boundary Value Problems for Force-Free Magnetic Fields},
	journal = {The Astrophysical Journal}
}

@article{BAYONAIII,
    title = {On the role of polynomials in RBF-FD approximations: III. Behavior near domain boundaries},
    author = {V. Bayona and N. Flyer and B. Fornberg},
    journal = {Journal of Computational Physics},
    volume = {380},
    pages = {378--399},
    year = {2019},
    doi = {https://doi.org/10.1016/j.jcp.2018.12.013}
}

@article{Manchester04,
    author = {Manchester IV, W. B. and Gombosi, T. I. and Roussev, I. and Ridley, A. and De Zeeuw, D. L. and Sokolov, I. V. and Powell, K. G. and Tóth, G.},
    title = {Modeling a space weather event from the Sun to the Earth: CME generation and interplanetary propagation},
    journal = {Journal of Geophysical Research: Space Physics},
    volume = {109},
    number = {A2},
    pages = {},
    doi = {https://doi.org/10.1029/2003JA010150},
    year = {2004}
}

@article{FE14,
    author = {Fornberg, B. and Elcrat, A. R.},
    title = {Some observations regarding steady laminar flows past bluff bodies},
    journal = {Phil. Trans. Royal Society A},
    volume = {372},
    number = {20130353},
    pages = {},
    doi = {http://dx.doi.org/10.1098/rsta.2013.0353},
    year = {2014}
}

@article{lsqr,
    author = {Paige, C. C. and Saunders, M. A.},
    title = {LSQR: An Algorithm for Sparse Linear Equations and Sparse Least Squares},
    journal = {ACM Trans. Math. Software},
    volume = {8},
    number = {1},
    pages = {43--71},
    year = {1982}
}

@article{torok05,
	doi = {10.1086/462412},
	year = 2005,
	month = {8},
	volume = {630},
	number = {1},
	pages = {L97--L100},
	author = {T. T\"{o}r\"{o}k and B. Kliem},
	title = {Confined and Ejective Eruptions of Kink-unstable Flux Ropes},
	journal = {The Astrophysical Journal}
}

@article{dove11,
       author = {J. B. Dove and S. E. Gibson and L. A. Rachmeler and A. Tomczyk and P. Judge},
        title = {A Ring of Polarized Light: Evidence for Twisted Coronal Magnetism in Cavities},
      journal = {The Astrophysical Journal Letters},
         year = 2011,
        month = 4,
       volume = {731},
       number = {1},
          eid = {L1},
        pages = {L1},
          doi = {10.1088/2041-8205/731/1/L1}
}

@article{gibson08,
    author = {Gibson, S. E. and Fan, Y.},
    title = {Partially ejected flux ropes: Implications for interplanetary coronal mass ejections},
    journal = {Journal of Geophysical Research: Space Physics},
    volume = {113},
    number = {A9},
    pages = {},
    doi = {https://doi.org/10.1029/2008JA013151},
    year = {2008}
}

@article{malanushenko20,
    author = {A. Malanushenko and N. Flyer and S. Gibson},
    title = {Convolutional neural networks for predicting the strength of near-earth magnetic field caused by interplanetary coronal mass ejections},
    journal = {Frontiers in Astronomy and Space Sciences},
    volume = 7, 
    pages = {62}, 
    doi = {10.3389/fspas.2020.00062},
    year = {2020}
}

@article{lugaz13,
    author = {N. Lugaz and C. J. Farruglia and W. B. Manchester IV and N. Schwadron},
    title = {The interaction of two coronal mass ejections: influence of relative orientation},
    journal = {The Astrophysical Journal},
    volume = {778},
    number = 1,
    pages = {20},
    year = {2013}
}

@article{manchester03,
    author = {Manchester IV, W. B. and Vourlidas, A. and T\'{o}th, G. and Lugaz, N.and Roussev, I. I. and Sokolov, I. V. and Gombosi, T. I. and de Zeeuw, D. L. and Opher, M.},
    title = {Three-dimensional MHD Simulation of the 2003 October 28 Coronal Mass Ejection: Comparison with LASCO Coronograph Observations},
    journal = {The Astrophysical Journal},
    volume = {684},
    number = 2,
    pages = {1448-1460},
    doi = {10.1086/590231},
    year = {2003}
}

@article{gibson00,
    author = {S. E. Gibson and B. C. Low},
    year = {2000},
    title = {Three-Dimensional and Twisted: An MHD interpretation of on-disk characteristics of coronal mass ejections},
    journal = {Journal of Geophysical Research},
    volume = 105,
    number = {A8},
    pages = {18187-18202},
    doi = {10.1029/1999JA000317}
}

@article{nindos20,
    author = {A. Nindos and S. Patsourakos and A. Vourlidas and X. Cheng and J. Zhang},
    year = {2020},
    title = {When do solar erupting hot magnetic flux ropes form?},
    journal = {Astronomy \& Astrophysics},
    volume = {642},
    pages = {A109},
    doi = {10.1051/0004-6361/202038832}
}

\appendix
\section{Gibson-Low Model}\label{Appendix_GibLow}

\textcite{gibson98} provides a full, time-dependent MHD solution by a self-similar polytropic expansion of an MHS equilibrium. Just the MHS equilibrium is described below.

First note that a solution to the MHS equations as given in \eqref{eq_mhs} is equivalent to solving the system without gravity in spherical coordinates centered at the center of the sun as
\begin{equation}\label{eq_appendix_nodensMHS}
\begin{array}{c}
    \frac{1}{4\pi} (\nabla\times\mathbf{b})\times\mathbf{b}-\nabla\Pi = 0\\
    \nabla\cdot\mathbf{b}=0
\end{array}
\end{equation}
and then deforming the solution with a radial stretching $r\rightarrow r+a$ with $a$ constant. As such, a solution will be first presented in terms of $\mathbf{b}$ and $\Pi$, and then modified to include gravitational effects.

Finally, the spherical construction can be converted to cartesian coordinates in the typical way. The construction provided below is as it is done in \textcite{gibson98}, wherein $\hat{x}$ is the axis perpendicular to the photosphere. In this work, the coordinates are then relabeled so that $\hat{z}$ is the vertical axis.

\subsection{General Solution Construction}

The domain of the model is split into two regions by a roughly spherical domain $\sigma$ of radius $r_0$ placed a distance $r_1$ from the center of the Sun. Inside the sphere the magnetic field lines are closed, looping back in on themselves. Outside the sphere the magnetic field lines are open, trailing off to infinity. It is useful to define two coordinate systems: $(x,y,z)$ (Cartesian) or $(r,\theta,\phi)$ (Spherical) centered at the center of the sun with $r=1$ the photospheric surface, and $(x',y',z')$ and $(r',\theta',\phi')$ centered at the center of $\sigma$. The heliocentric coordinate system is defined such that $\sigma$ is placed along the $x$ axis. Following the original model paper, the spherical coordinate system typical to physics is used here, with $\theta$ denoting the polar angle.

Define the magnetic field outside the sphere as $\mathbf{b}_{ext}$ and the magnetic field inside the sphere as $\mathbf{b}_{int}$. As $\mathbf{b}_{ext}$ is symmetric about the $z$ axis, it may be written in terms of a stream function $A_{ext}$, as in
\begin{equation}
    \mathbf{b}_{ext} = \frac{1}{r\sin(\theta)}\left(\frac{1}{r}\frac{\partial A_{ext}}{\partial\theta}\hat{r} - \frac{\partial A_{ext}}{\partial r}\hat{\theta}\right).
\end{equation}
The interior field $\mathbf{b}_{int}$ is axisymmetric about $x'$ and can so similarly be written in terms of the stream function $A_{int}$ as
\begin{equation}
    \mathbf{b}_{int} = \frac{1}{r'\sin(\theta')}\left(\frac{1}{r'}\frac{\partial A_{int}}{\partial\theta'}\hat{r}'-\frac{\partial A_{int}}{\partial r'}\hat{\theta}' + \alpha_0 A \hat{\phi}'\right)
\end{equation}
where the Bessel function $J_{5/2}(\alpha_0r_0)=0$, i.e. $\alpha_0 = 5.763854/r_0$.

The final solution is then constructed as
\begin{equation}
\begin{split}
    A_{ext} &= \underbrace{\cos(\theta)}_{\Psi_0} + 
         \underbrace{-\frac{1}{r_0}\frac{r_1(r^2+r_1^2-r_0^2)+r\cos(\theta)(r_0^2-2r_1^2)}{\left[(r_0^2-r_1^2)^2 + r_1^2r^2 + 2rr_1(r_0^2-r_1^2)\cos(\theta)\right]^{1/2}}}_{\Psi_1} + 
         \underbrace{\frac{1}{r_0}\left[r^2+r_1^2-2rr_1\cos(\theta)\right]^{1/2}}_{\Psi_2}\\
    A_{int} &= \frac{4\pi a_1}{\alpha_0^2}\left[\frac{\sin(\alpha_0 r')/(\alpha_0 r') - \cos(\alpha_0 r')}{\sin(\alpha_0 r_0)/(\alpha_0 r_0) - \cos(\alpha_0 r_0)}r_0^2-r'^2\right]\sin^2(\theta')\\
    \Pi_{int} &= a_1 A_{int}
\end{split}
\end{equation}
with $a_1$ a parameter detailed in the original model paper. The final magnetic field and pressure that solves \eqref{eq_appendix_nodensMHS} is then dictated by $\mathbf{b}_{int}$ inside $\sigma$ with a discontinuous transition to $\mathbf{b}_{ext}$ outside $\sigma$. The sign of $b_{r,ext}$ is further flipped over the solar equator.

\subsection{Radial Deformation}

Replace the variable $r$ in $\mathbf{b}(r,\theta,\phi)$ with $\Lambda(r)=r+a$ with $a$ a parameter. Then with $G$ the gravitational constant and $M_\odot$ the solar mass,
\begin{equation}
\begin{split}
    B_r(r,\theta,\phi) =& \left(\frac{r+a}{r}\right)^2b_r(r+a,\theta,\phi)\\
    B_\theta(r,\theta,\phi) =& \frac{r+a}{r}b_\theta(r+a,\theta,\phi)\\
    B_\phi(r,\theta,\phi) =& \frac{r+a}{r}b_\phi(r+a,\theta,\phi)\\
    P =& \frac{(r+a)^2}{r^2}\left(1-\frac{(r+a)^2}{r^2}\right)\frac{b_r^2}{8\pi} + \frac{(r+a)^2}{r^2}\Pi\\
    \rho =& \frac{r^2}{GM_\odot}\Bigg[-\frac{(r+a)^2}{r^2}\left(1-\frac{(r+a)^2}{r^2}\right)\frac{d}{d\Lambda}\left(\Pi+\frac{||\mathbf{b}||^2}{8\pi}\right) + 2\frac{(r+a)a}{r^3}\Pi +\\
        & \frac{(r+a)a}{4\pi r^3}\left(1-2\frac{(r+a)^2}{r^2}\right)b_r^2 + \frac{(r+a)^2}{r^2}\left(\frac{a^2}{r^2}+\frac{2a}{r}\right)\left(\frac{b_\theta^2+b_\phi^2}{4\pi(r+a)}\right)\Bigg].
\end{split}
\label{eq_GLsoln}
\end{equation}

\end{document}